\input{aipcheck}
\documentclass[
    ,final            
  ]
  {aipproc}
\layoutstyle{8x11double}
\begin{document}
\title{Chaotic behavior of micro quasar GRS~1915+105}
\author{Banibrata Mukhopadhyay}{address={Astronomy Division, University of Oulu, 
P.O.Box 3000, FIN-90014, Finland}}

\begin{abstract}

Black hole binaries are variable in timescales of rang from months to milli-seconds.
The origin of this variability is still not clear, it could be due to the variation of 
external parameters, like mass accretion rate, instabilities in the inner regions of
the accretion flow etc. Important constraints on these possibilities can be obtained
from the study of the non-linear behavior of fluctuations. We present a modified non-linear
time series analysis technique which optimizes the use of the available data and
computes the correlation dimension in a non-subjective manner. We apply this
technique to the X-ray light-curve of the black hole system, GRS~1915+105, to 
show conclusively that at least for four of its twelve temporal classes, the underlying 
mechanism is a low order chaotic one.

\end{abstract}

\maketitle
\section{1. Introduction}

One of the most interesting black hole candidates observed so far is
GRS~1915+105. Like other black hole sources, it shows the X-ray variability
in a wide range of timescale varies from months to milli-seconds, 
which certainly indicates that the system is highly non-linear (that is even
true for other black holes). 
However, the most exciting thing behind this micro quasar GRS~1915+105 
is that, according to the temporal variability it can be classified into twelve 
different temporal classes with respect to the morphology of 
light-curves for different observation IDs (OIDs) \cite{bel00}.
But, from this temporal classification
one is unable to understand about a basic feature of non-linearity, 
whether this black hole is a random or chaotic system,
that is our present goal to understand. Following an established technique of
non-linear dynamical physics applied earlier to other related astrophysical
problems \cite{unno90,tim00,th01}, here we plan to establish that the 
micro quasar GRS~1915+105 as well as the black system is chaotic in nature.         
This analysis of temporal behavior of a system plays an important role  
to understand the geometry of the source, which thus 
eventually be used to test its relativistic nature and the corresponding accretion
process.  

Before going into detail of our analysis, let us introduce some basic definitions
of various terminology used in this article.
\begin{description}
\item[Chaos:]
 If any two consecutive trajectories of a system, while divergent
each other, are related by some law, the system is called chaotic
(eventhough instantaneously it looks like random but overall it
is deterministic). Brownian motion is an immediate practical example of
it. In case of the accretion disk physics, one can check this by 
investigating various orbits around a compact object.  
\item[Random:] 
If any two consecutive trajectories of a system are not related by any
physics, the system is called random. Poisson noise is an example.
\item[Degrees of Freedom:]
From the concept of classical mechanics, the {\it Free Degrees of Freedom}
of an $M$ dimensional system is $M$. If the number of constraints into
the system is $n$, the {\it Net Degrees of Freedom} of the same system 
can be defined as, $D=M-n$. Philosophically, same thing is true even in
the case of non-linear dynamics, however, the estimated $D$ need not
be integer (unlike any classical mechanical system). In non-linear
dynamics, $D$ is the measure of chaos, called {\it chaotic dimension} of the system.
If there is no constraint into the system at any dimension, $M$; $D=M$, both the degrees of
freedom are same, and therefore the system is random. Naturally, for a constraintless
random system $D$ varies linearly with $M$ while for a chaotic system
$D$ saturates to a value above the particular $M$. For an ideal chaotic system,
$D$ saturates when $M\ge2$, thus sometimes denoted as $D_2$.    
\end{description}

The chaotic nature of accretion phenomena in magneto-hydrodynamic simulations
has been found by Winters et al. \cite{win03} already. Therefore we 
are motivated to check this chaotic nature from the observational point of
view, analyzing black holes data. If the fluctuations in an accretion disk is random
or stochastic, the corresponding X-ray variations are expected due to the
variation of external parameters, e.g. accretion rate, and/or there is 
a possibility of random flares and vice versa. In contrary, the presence
of chaotic nature, which is deterministic, may be due to the inner disk
instability and/or coherent flares and vice versa.

To determine the chaotic nature of GRS~1915+105, in the next section, we 
outline the methodology and in \S 3 discuss results for different OIDs. Finally in
\S 4, we make our conclusions.

\section{2. Method}

Once we have the light-curve data, let us denote the count rate at the
time $t_j$ be $s(t_j)$. Let us also identify a {\it delay}, $\tau$, that indicates
the time interval beyond which any two count rates are not related. This is the time
at which the auto-correlation function of the system goes to zero or reaches
to its first minima (if it does not go to zero). If the auto-correlation function 
neither goes to zero nor attains a minimum, the mentioned {\it delay time} has to be 
figured out in a different way explained in \S 3. Therefore, using this delay one can
construct a vector at time $t_j$ in an $M$ dimensional space as
\begin{equation}
\vec{x}_j=\vec{x}(t_j)=\{s(t_j),s(t_j+\tau),s(t_j+2\tau),......{\rm up\hskip0.1cm to
\hskip0.1cm}M^{th}{\rm \hskip0.1cm s}\}.
\label{vec}
\end{equation}
In this way all the possible $M$ dimensional vectors have to be constructed for
$j=1,2,3,...,N$, and one can get an $N\times M$ matrix equation.
Then one has to map all the vectors in an $M$ dimensional phase-portrait, and
to compute the average number of data points within a distance $R$ from a particular
data point\footnote{The average number of data points in an $M$-cube of arm length $R$ 
or $M$-sphere of radius $R$ 
in the phase-portrait.}. For example, in case of $M=2$, one has to find out that
mentioned average number of points in the $s(j+\tau)-s(j)$ space. 
This is called the correlation sum, defined as 
\begin{equation}
C_M(R)=\lim_{N\rightarrow \infty}\frac{1}{N(N-1)}\sum_i^N \sum_{j\neq i}^N 
H(R-|{\vec x}_i-{\vec x}_j|),
\label{cor}
\end{equation}
where $H(R-|{\vec x}_i-{\vec x}_j|)$ is a Heaviside step function.
Then one has to find out $C_M(R)$ for different $R$s. Finally,
the whole process has to repeat for the various choices of $M$ (say $1\rightarrow 15$).
Subsequently, the variation of $log[C_M(R)]$ as a function of $log[R]$ 
when $M$ is a parameter, has to be seen concentrating on the (approximate)
linear region of curves. Finally the average slope of approximate linear 
region of the curve for different $M$ to be computed as the correlation dimension of the 
system, defined as
\begin{equation}
D_2=\frac{d[logC_M(R)]}{d[log(R)]}.
\label{cordim}
\end{equation}
Now from the variation of $D_2$ as a function of $M$, one can understand whether
the system is chaotic or random as mentioned in \S 1, and the chaotic dimension can be 
identified from the saturated value of $D_2$ in the curve. The details of all these
will be presented elsewhere \cite{mis-prep}.    

An important point to be noted here that for a very small $R$ ($\le R_{min}$), $C_M(R)$ would be
of the order of unity and the result would be Poisson noise dominated.
On the other hand, for a large $R$ ($\ge R_{max}$), $C_M(R)$ would saturate to 
the total number of 
data points. Therefore for a particular $M$, there is a range of $R$ which gives
the physical result where the $C_M(R)-R$ curve is linear. Also the maximum
$M$ (largest $M$-cube/sphere) is chosen in such a manner that it has to be within the 
embedding space so that filled by points and there should not be any edge effect
due to the limitation of point's number. Due to all these mentioned reasons,
if $M$ is above of a particular value $M_c$ ($M>M_c$) there is a chance that $R_{min}=R_{max}$,
and then no significant results can be obtained.

Now we like to apply all the above mentioned technique to the data of
GRS~1915+105 to understand whether it behaves as chaos or random. In the next
section, we discuss this.

\section{3. Results}

We take the RXTE data of OIDs corresponding to each of the temporal classes of GRS~1915+105. 
It is found that
different OIDs for a particular class give same results upto an error bar. Therefore,
we choose one OID for each temporal class to present our results, 
given in Table \ref{tab}. For each class we extract a few continuous data streams
$\sim 3000$ sec long. The time resolution of light-curves is chosen as
$0.1$ sec, which gives $\sim 30000$ data points for each of them with $\sim 1000$ counts
per bin. Light-curves for a finer time resolution are Poisson noise dominated,
while with a larger binning give a very little number of data points for the
present purpose. In our cases, the auto-correlation function neither goes to
zero nor attains a minimum, therefore that $\tau$ is chosen in our
calculations above which the $D_2-M$ curves saturate. Here this saturated $\tau$
is typically $\sim 20-50$, and we choose it as $\tau=50$ for all the cases.

\begin{figure}
  \includegraphics[height=.3\textheight]{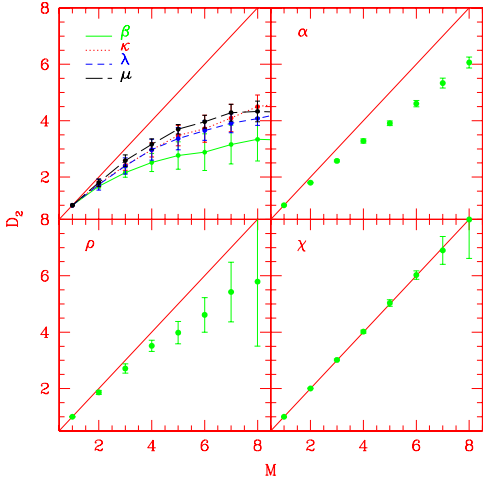}
\label{fig1}
  \caption{Results for GRS~1915+105 data in seven temporal classes. The curves for 
$\kappa$, $\mu$, $\beta$ and $\lambda$ classes indicate chaotic signature, while that for 
$\chi$ shows random or stochastic nature of the system. The cases for $\alpha$ and $\rho$ 
indicate {\it some deviation} from random signature.} 
\end{figure}

In Fig. \ref{fig1}, we show results for seven temporal classes of
GRS~1915+105 data following the method outlined in \S 2. According to the description
of chaos, random, and degrees of freedom in Introduction, the solid diagonal line 
in each of the boxes in the figure indicates the {\it ideal random curve}, $D_2=M$ line,
when there is no constraint into the system. Figure \ref{fig1} shows
the results of $\beta$, $\kappa$,
$\lambda$ and $\mu$ cases those indicate a clear deviation of this ideal random curve
which depict a signature of chaos of dimension $\sim 3.3-4.5$. In contrary, the curve of
$\chi$ case perfectly overlaps with $D_2=M$ line which indicates an ideal random signature.   
Also the temporal classes $\gamma$ and $\phi$ show a similar random signature.
The cases of $\alpha$ and $\rho$ show a deviation of the ideal random, but that is not
as much as of chaos cases. Therefore we call this kind of situations as semi-random 
(or semi-stochastic). Similar semi-random behaviors come out from the analysis of
$\theta$, $\nu$ and $\delta$ temporal classes. 

\begin{figure}
  \includegraphics[height=.3\textheight]{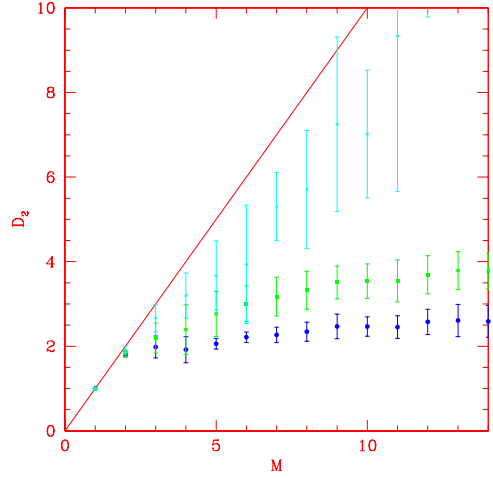}
\label{fig2}
 \caption{The effect of Poisson noise to the Lorenz data. The curve with circles indicates
the result for actual Lorenz data. The curves with squares and triangles come out 
when the Lorenz data is rescaled by Poisson noise to $\beta$ and $\gamma$ like classes
(with the same average count and rms variation).
}
\end{figure}

At this point, we could divide the results as well as GRS~1915+105 system into three
different classes or stages: low-dimensional chaos (deterministic), 
random or stochastic (indeterministic) and semi-random, as far as the non-linear dynamical
analysis is concern. However, we like to perform a {\it test} before making any strong 
statement. We know that the Lorenz system is a model of low dimensional chaos where
the chaotic dimension, $D_2=2.04$. That means for the data of Lorenz system, $D_2-M$
curve starts to saturate when $M=2$ with the saturation $D_2$ is $2.04$. 
On the other hand the $D_2-M$ curve for the Poisson noise appears as a straight line 
of unit gradient passing through the origin (i.e. $D_2=M$). Let us take the
Lorenz data set and introduce Poisson noise into it in such a manner that
the average count and rms variation become same as that of $\beta$ case. Then using this 
modified {\it noise induced Lorenz data}, if we perform $D_2-M$ analysis again, now the surprising
fact comes out that $D_2$ no longer saturates to $2.04$, instead becomes increased
to $\sim 4$. If a higher order rescaling (as mentioned above)
to Lorenz data is performed, such that   
the average count and rms variation are same as that of $\gamma$ case, the 
$D_2$ becomes more and more increased, approached toward the $D_2=M$ line though not
exactly overlapped on it. All these scenarios have been pictorially represented in 
Fig. \ref{fig2}. From these discussions, it is very clear that presence of any
kinds of noise converts any low dimensional chaotic system to that of high dimensional
and/or random or semi-random. It does not matter whether the system has any chaotic 
signature or not, noise always suppress it and the system practically 
appears as higher dimensional or random. Therefore the computed chaotic
dimension $\sim 4$ (comes out from Fig. \ref{fig1}) 
is an over estimation. If the noise would have been
possible to remove from the system, those could appear as a low dimensional chaos
like Lorenz system. Similarly, the random and semi-random appearance of, say, $\chi$ and 
$\alpha$ classes respectively, are only due to the dominance of noise into the system.
If the system would have been noise free (or less noisy) those classes could also
be appeared as chaos or/and low dimensional chaos. This dominance of noise into
the system in the random cases will be more clear if we look on to the Table \ref{tab},
which clearly shows the Poisson noise to rms ratio is higher for the random cases compared
to that for the cases of chaos. The table also lists the various OIDs corresponding to the temporal
classes, the average count, rms variation, Poisson noise and finally what
the analysis tells about it, whether the system is chaos: C, random/stochastic: S or 
semi-random/stochastic: SS. 

\begin{table}
\begin{tabular}{lrrrrrrr}
\hline
\hline
 & \tablehead{1}{l}{b}{OID} 
 & \tablehead{1}{r}{b}{Class} 
 & \tablehead{1}{r}{b}{$<S>$}
 & \tablehead{1}{r}{b}{rms}
 & \tablehead{1}{r}{b}{$<PN>$}
 & \tablehead{1}{r}{b}{$<PN>$/rms}
 & \tablehead{1}{r}{b}{Stage} \\
\hline
\hline
& 10408-01-10-00 & $\beta$  &1917 &  1016 & 43.8 &0.04& C \\
& 20402-01-37-01 & $\lambda$ & 1493 & 1015 & 38.6 & 0.04 & C \\
& 20402-01-33-00 & $\kappa$ &   1311 & 800 & 36.2 & 0.04 & C \\
& 10408-01-08-00 & $\mu$ &         3026 & 999 & 55 & 0.06 & C \\
\hline
& 20402-01-45-02 &  $\theta$ &  1740 &  678 & 41.7 & 0.06 & SS \\
& 10408-01-40-00 & $\nu$    &  1360 & 462   & 36.9 & 0.08 & SS \\
& 20402-01-03-00 & $\rho$   &        1258 &  440  & 35.5 & 0.08 & SS \\
& 20187-02-01-00 & $\alpha$ &   582 &  244  & 24.1 & 0.10 & SS \\
& 10408-01-17-00 &  $\delta$ &      1397 &  377  & 37.4 & 0.10 & SS \\
\hline
& 20402-01-56-00 &  $\gamma$ & 1848 &  185 &  43.0 & 0.23 & S \\
& 10408-01-22-00 &  $\chi$ & 981  & 118 & 31.3 & 0.27 & S \\
& 10408-01-12-00 &  $\phi$ & 1073 & 118  & 32.7 & 0.28 & S \\
\hline
\hline
\end{tabular}
\caption{Columns:- 1. RXTE OID, 2. Temporal class of the system 
according to \cite{bel00},
3. Average count in the light-curve, $<S>$, 4. Root mean square variation
in the light-curve, $rms$, 5. Expected Poisson noise variation, $<PN> \equiv \sqrt{<S>}$,
6. Ratio of the expected Poisson noise to root mean
square variation, 7. Stage of the system as understood from $D_2-M$ curves 
(C: chaotic; SS: semi-stochastic; S: stochastic)  }
\label{tab}
\end{table}

\section{4. Conclusions}

We analyze the non-linear behavior of the micro quasar GRS~1915+105 in 
terms of the signature of chaos and random. It immediately comes out that
at least four out of its twelve temporal classes are chaotic in nature.   
Therefore, there is no doubt that GRS~1915+105 behaves as chaos at least in some stages. 
The three out of those remaining eight classes depict as random
while five others show a deviation from random, called as semi-random.
By a simple test, i.e. introducing noise into the low dimensional
chaotic system, and from the ratio of Poisson noise to rms variation
in the data of various classes, it comes out that the random and semi-random
cases are noise dominated. Therefore, we can hypothesize that GRS~1915+105
is chaotic in nature. This chaotic signature is suppressed only in some of its stages 
due to the noise dominance and it appears like random or semi-random, but actually it is not that.

In early, the results of Cyg X-1 data seemed to be random or very high dimension
chaos \cite{unno90}. On the other hand the temporal behavior of Cyg X-1 is very 
similar to that of the $\chi$ class of GRS~1915+105, which is noise dominated depicted
as random according to our analysis. Therefore, we understand that due to the
dominance of noise into the system, Cyg X-1 could not show its chaotic nature,
what it could be actually. In an alternative way, we can say that there may
be a stochastic component to the variability which dominates for certain
temporal states. 

Finally, we can conclude that any black hole system may be chaotic in nature.
Depending on the order of noise present into the system, it appears either as actual
chaos or random. Any random or semi-random nature may not be its fundamental
signature. If the noise would have been possible to remove from the system, always
it could show an actual chaotic signature. Overall, the identification
of chaotic nature of black hole systems has opened a new window to understand
their temporal behavior deeply. In order to have a more concrete knowledge and to upgrade the confidence
level, the next step should be to study the corresponding Lyapunov exponent
(which is another basic measure of a non-linear system to distinguish the chaos from
random) and the associated Kolmogorov entropy. 

\begin{theacknowledgments}
Author would like to thank R. Misra, K. P. Harikrishnan, G. Ambika and A. K. Kembhavi
for useful discussions and for collaboration on a project of which
the work reported here is a part. Thanks are also given to J. Poutanen for useful
discussions.
The partial support by Academy of Finland grant 80750 to this work is acknowledged.
\end{theacknowledgments}


\bibliographystyle{aipproc}   
\bibliography{proc}

\hyphenation{Post-Script Sprin-ger}
\begin{thebibliography}{9}
\expandafter\ifx\csname natexlab\endcsname\relax\def\natexlab#1{#1}\fi
\providecommand{\enquote}[1]{``#1''}
\expandafter\ifx\csname url\endcsname\relax
  \def\url#1{\texttt{#1}}\fi
\expandafter\ifx\csname urlprefix\endcsname\relax\def\urlprefix{URL }\fi

\bibitem[1]{bel00}
Belloni, T., Klein-Wolt, M., M\'endez, M., van der Klis, M., and van Paradijs, J., 
\emph{A\&A}, \textbf{355}, 271--290 (2000).

\bibitem[2]{unno90}
Unno, W., Yoneyama, T., Urata, K., Masaki, I., Kondo, M. and Inoue, H.,
\emph{PASJ}, \textbf{42}, 269--278 (1990).

\bibitem[3]{tim00}
Timmer, J., Schwarz, U., Voss, H., Wardinski, I., Belloni, T.,
Hasinger, G., van der Klis, M. and Kurths, J.,
\emph{PRE}, \textbf{61}, 1342--1352 (2000).

\bibitem[4]{th01}
Thiel, M., Romano, M., Schwarz, U., Kurths, J., Hasinger, G. and Belloni, T.,
\emph{ApSSS}, \textbf{276}, 187--188 (2001).

\bibitem[5]{win03}
Winters, W., Balbus, S. and Hawley, J.,
\emph{MNRAS}, \textbf{340}, 519--524 (2003).

\bibitem[6]{mis-prep}
Misra, R., Harikrishnan, K., Mukhopadhyay, B., Ambika, G. and Kembhavi, A., \emph{ApJ},
\textbf{to appear} (2004).

%
%
%
%
%
%

\end{thebibliography}
\end{document}